\documentclass[12pt]{indmath}

\usepackage{neubauer}
\usepackage{graphicx}
\usepackage{mathptmx} 
\usepackage{amsmath}
\usepackage{xcolor}
\usepackage{caption}
\usepackage{soul}
\usepackage{subcaption}
\def\Re{\mfunc{Re}}

\begin{document}

\title{A generalised non-linear reconstructor for all Fourier-type wavefront sensors}
\author{Victoria Laidlaw$^1$\\\footnotesize{\rm $^1$Industrial
 Mathematics Institute, Johannes Kepler University Linz, A-4040 Linz,
 Austria.
\\ Corresponding author: victoria.laidlaw@indmath.uni-linz.ac.at}}
\maketitle


\begin{abstract}
State-of-the-art adaptive optics (AO) systems perform non-linear Fourier-type wavefront sensing for real-time corrections of dynamic wavefront aberrations. This general class of sensors uses a filtering mask in the focal plane that converts phase fluctuations of the incoming light into intensity variations in the subsequent pupil plane. Due to their high sensitivity, Fourier-type wavefront sensors (WFSs) are the sensors of choice for many current and upcoming AO systems in ophthalmic imaging, free-space optical communications (FSOC) and astronomical ground-based telescopes such as the forthcoming generation of extremely large telescopes (ELTs). Conventionally, linear methods, like a matrix-vector-multiplication (MVM), are used for the inversion of Fourier-type WFSs. However, their non-linear behavior gives rise to severe performance degradations when significant channel perturbations are observed. They are expected to occur during relatively strong atmospheric turbulence conditions, which are commonplace for both non-rural sites and daytime observations. Other sources for these conditions can be non-common path aberrations and short sensing wavelengths.

This study presents a novel type of iterative reconstructor to overcome non-linear wavefront sensing regimes. The underlying method is the non-linear Landweber iteration with Nesterov acceleration, well known in the field of inverse problems. A significant advantage of the new approach is its direct applicability to any Fourier-type WFS. This is implemented by adapting the filtering mask of the specific Fourier-type WFSs in the model-based algorithm.

Several Fourier-type wavefront sensors are considered for ELT-scale instruments and their performance with the new algorithm is compared. The study goes on to concentrate on the pyramid wavefront sensor (PWFS), one of the most well-known Fourier-type WFSs. We demonstrate in end-to-end simulations that this novel approach outperforms linear methods in non-linear sensing regimes.

\end{abstract}

\section{Introduction}\label{sec_intro}
When optical signals from, e.g. stars or satellite laser terminals, are observed with ground-based telescopes, distortions of the incoming light are caused by it propagating through atmospheric turbulence~\cite{Roddier,RoWe96}. A similar effect is seen during retinal imaging for medical diagnostics. Distortions of the laser beam emerge from imperfect light propagation through the cornea, lens and vitreous body of the eye~\cite{Liang97}. Advanced adaptive optics (AO) systems aim to correct the effects of dynamic wavefront aberrations in real-time. They typically do this by employing a number of wavefront sensors (WFSs), which use a control algorithm to convert their measurements into commands to drive a number of deformable mirrors (DMs). Instruments using this technology are implemented in ground-based telescopes, such as those used for free-space optical communications (FSOC) as well as the forthcoming generation of extremely large telescopes (ELTs). They also play a crucial role in ophthalmology for the early detection of abnormalities and diseases.

\bigskip \par

\begin{figure}
  \centering
  \includegraphics[width=0.9\textwidth]{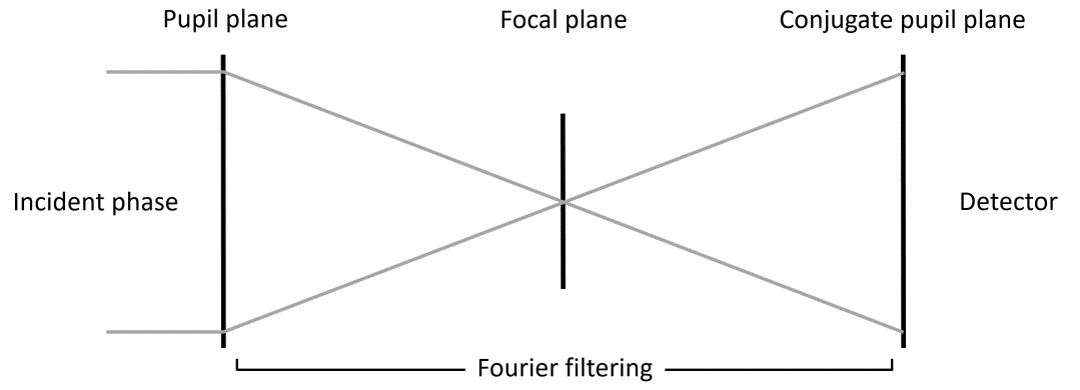}
  \caption{Sketch of a Fourier-type WFS~\cite{HuNeuSha_2023}.}
  \label{fig_fwfs}
\end{figure}

Non-linear Fourier-type WFSs~\cite{Fauv16} have become particularly interesting for measuring wavefronts in AO due to their high sensitivity. They use optical Fourier filtering with a suitable optical element (e.g., a multi-facet glass prism) located in the focal plane (see fig.~\ref{fig_fwfs}). The optical element splits the electromagnetic field
into several beams, each of which produces a differently filtered image of the
entrance pupil in the subsequent pupil plane. The intensity patterns are then measured by a camera. The
underlying mathematical model for the intensity $I$ is given by
\be{i1}
 (I(\phi))(x,y)=\abs{F^{-1}\kl{OTF\cdot F\kl{\chi_\Omega e^{-i\phi}}}(x,y)}^2
\ee
for $F$ representing the Fourier transform, $\phi$ the incoming wavefront and $\chi_\Omega$ the characteristic function of the pupil $\Omega$. The optical transfer function $OTF$
describes the optical element that divides the light in the focal plane. Please note that in equ.~\req{i1} we consider a simplified version of the actual incoming electromagnetic field $$\sqrt{n}\chi_\Omega e^{-i\phi},$$ where $n$ is the spatial average incoming flux. As $n$ is proportional to the total flux on the detector, this normalisation can always be done in post-processing. Furthermore, it is assumed that the incoming light is monochromatic and $$\phi = \frac{2\pi}{\lambda}\Delta$$ is the perturbed phase at the considered wavelength $\lambda$. The optical path difference $\Delta$ was created by atmospheric turbulence or another source of perturbation.

A popular Fourier-type WFS is the so-called pyramid wavefront sensor (PWFS)~\cite{Raga96}. It is included in the design of many current and upcoming astronomical AO systems. It is the primary WFS in several first-light instruments of the ELT and is also considered by the European Space Agency (ESA) for ground-space FSOC~\cite{alasca2023}. In addition, the PWFS has recently been shown to provide high-performance retinal imaging in vivo and is currently undergoing clinical studies for patients suffering from diabetic retinopathy~\cite{Brunner21,ARVO_2024}.

The optical element in the focal plane of the PWFS is a 4-sided pyramidal glass prism. It splits the electromagnetic field into $4$ beams. The corresponding transfer function of the prism is
\begin{equation*}\label{otf} 
 OTF_{4pyr}(\xi,\eta)=e^{ic(|\xi|+|\eta|)}\ .
\end{equation*}

\begin{figure}
  \centering
  \includegraphics[width=0.99\textwidth]{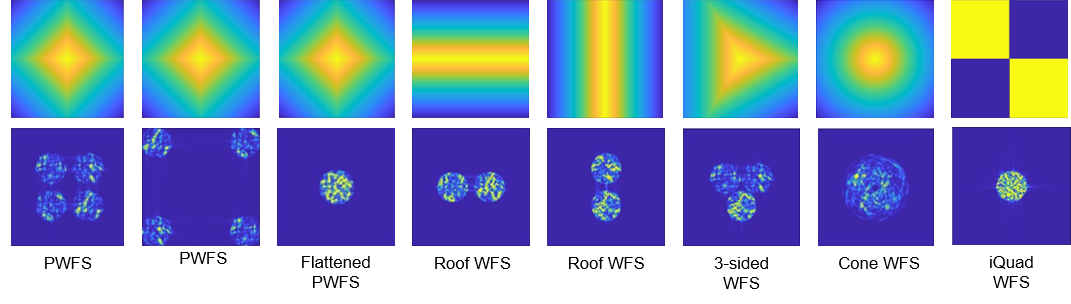}
  \caption{Transfer function of different Fourier-type WFSs (top) and corresponding detector intensities for an example wavefront (bottom). The first three columns represent a PWFS but with different apex angles. The third column is a flattened PWFS where all four intensity pupil images are overlapping.}
  \label{fig_otf}
\end{figure}

The constant $c>0$ refers to the angle of apex of the pyramidal prism, which influences the separation distance between the four intensity patterns in the conjugate pupil plane. The PWFS $OTF$ and its corresponding intensity pattern for an example wavefront are visualised in fig.~\ref{fig_otf} where the first three columns represent PWFSs with different angles $c$. The WFSs can make a measurement by considering only the light on the pupils or the full frame, i.e., taking into account the interference effects and the light between the pupils. Although the full-frame approach can be more accurate than the pupils-only approach, its computational load can be challenging for ELT-scaled systems running in real-time~\cite{FaHuLAM_ao4elt6_poster}.

Fourier-type WFSs are a general class of WFSs in the sense that different optical elements can be used to split the light in the focal plane. Possible choices are $n$-sided pyramidal prisms for $n \in \N$ as is the case for the 2-sided roof WFS or the
3-sided pyramid WFS~\cite{Clare_ao4elt5,engler2017,Veri04}. The cone WFS is the extension of the idea to an infinite number of faces~\cite{ClareCone2020}. The Zernike WFS has a small circular hole with depth $\delta$ and diameter $p$ in the center of the mask of the optical element~\cite{Zernike1934} and the iQuad WFS shows a focal plane that is divided into 4-quadrants around the origin, i.e., it has a Cartesian structure~\cite{FaHuShaRaLAM19_AO4ELTproc}. Other Fourier-type WFSs are the bi-orthogonal Foucault knife-edge sensors (sharp and grey Bi-O edge WFSs). They follow the principal idea of roof WFSs and are therefore also motivated by the Foucault knife-edge test~\cite{Veri2024}. The optical transfer functions of the above WFSs are all described by 
\be{allotfs}
OTF\left(\xi,\eta\right) = e^{i\psi\left(\xi,\eta\right)}
\ee
where the corresponding shape functions $\psi$ are listed in tbl.~\ref{table_otf} and visualised in fig.~\ref{fig_otf} with their corresponding intensity images for an example wavefront.

\begin{table}[ht!]
\centering
    \begin{tabular}{|c|c|c|c|}
        \hline  
        & x-Roof WFS
        & 3-sided PWFS
        & Zernike WFS
        \\
		\hline 
        $\psi(\xi,\eta)=$ 
        & $c|\xi|$ 
        & $\begin{cases} 
            -2c\xi \,, & \frac{-\pi}{3}< \arctan{\frac{\xi}{\eta}} < \frac{\pi}{3} \\ c(\xi-\sqrt{3}\eta)\,, 
            & \frac{\pi}{3}< \arctan{\frac{\xi}{\eta}} < \pi \\ c(\xi+\sqrt{3}\eta) \,, & \text{otherwise}
            \end{cases}$
        & $\delta\chi_{\left[-p/2,p/2\right]}\left(\xi,\eta\right)$ 
        \\
        \hline 
        & y-Roof WFS
        & Cone WFS 
        & iQuad WFS 
        \\
		\hline 
        $\psi(\xi,\eta)=$ 
        & $c|\eta|$  
        & $c\sqrt{|\xi|^2+|\eta|^2}$ 
        &  $\begin{cases} 
            \frac{\pi}{2}\,, & \xi \eta < 0 \, \\ 0 \,, & \text{otherwise} 
            \end{cases}$ \\
        \hline 
	\end{tabular}
    \caption{Shape functions $\psi$ for the roof WFSs, the 3-sided PWFS, the cone WFS, the Zernike WFS and the iQuad WFS. The constant $c>0$ relates to the apex angle of the prisms. For the Zernike WFS typical parameters are $\delta=\pi/2$ and $p=1.06\lambda/D$ for the sensing wavelength $\lambda$ and the telescope diameter $D$.}
    \label{table_otf}
\end{table}

Artificial intelligence (AI) has also been shown to be applicable to the optimisation of Fourier-type WFSs. So-called deep PWFSs have been introduced, which use AI to train an optical layer to find an optimal passive diffractive element placed at a conjugated Fourier plane of the PWFS pyramidal prism~\cite{Guzman_24}. This adds an optical preconditioner to the standard PWFS setup to boost its performance. The $OTF$ of the deep PWFS is
$$OTF_{DPWFS}\left(\xi,\eta\right) = e^{ic(|\xi|+|\eta|)+d(\xi,\eta)}$$ 
for the 2d function $d$ describing the trained optical layer. The proposed deep optics methodology can be used for the design of completely new filtering masks, independent of pyramidal prisms. By training the function $\psi$ of equ.~\req{allotfs} new Fourier-type WFSs can be invented that better fit the demands of specific AO applications.

Several AO groups have implemented spatial light modulators (SLMs) on their optical benches to test and compare different realisations of Fourier-type WFSs ~\cite{Akondi2013,engler2018,Potiron2019}. With programmable phase delays a wide range of mask geometries can be generated. However, for on-sky application it is beneficial to manufacture a physical optical element.

Substituting the optical elements in the focal plane corresponds to replacing the $OTF$ in the underlying mathematical model~\req{i1}~\cite{Fauv16}. All possible realisations of physical elements and hence transfer functions build the general class of Fourier-type WFSs.

\bigskip\par

Optimal AO performance requires WFS measurements to be converted into exact DM actuator commands. It is commonplace for this conversion to be made using interaction matrix-based approaches~\cite{Roddier}. The optimal DM shape is reconstructed by a matrix-vector-multiplication (MVM). These methods can be applied independently of the implemented WFS. Hence, all Fourier-type WFSs can be controlled with the same idea. MVMs imply that the WFS responses scale linearly, i.e., the AO system is only robust when it is operated within the linear regime of the WFS. However, in non-linear regimes, interaction matrix-based approaches critically degrade image quality due to approximation errors. The severity of these errors strictly depends on the strength of the dynamic aberrations and the sensing wavelength. Any irregularities in the system, e.g., non-common path aberrations, intensify the errors. Optical gain (OG) methods have been proposed to help overcome these effects~\cite{cham20,deo_spie_2018,Korkiakoski_08}, some of which require additional hardware~\cite{Cham2021}.

In this work, we present a new iterative wavefront reconstructor in a general framework for all Fourier-type WFSs. We demonstrate that the application of the non-linear control algorithm can compensate for errors related to the sensors' non-linearity. It should be noted that this work builds upon a previous study~\cite{HuNeuSha_2023} in which the authors introduced a method based on the non-linear Landweber iteration. This study presents an acceleration of the algorithm and also shows simulated results of its performance when applied to an astronomical AO system. A key benefit of the new wavefront reconstruction algorithm is its direct applicability to any Fourier-type WFS. The only input required is the corresponding $OTF$.

\bigskip \par 
Section~\ref{sec:trade-off} investigates the non-linearity related to Fourier-type wavefront sensing and summarises existing methods to tackle non-linear approximation errors. In section~\ref{sec:nonlinear_solver} the idea of the non-linear iterative solver presented in ref.~\cite{HuNeuSha_2023} is recapped and an acceleration of the algorithm is presented. The previously developed theory is applied to several Fourier-type WFSs in section~\ref{sec:numsim} including end-to-end simulations of an ELT-scale instrument. Section~\ref{sec:conclusions} summarises the theoretical and numerical results presented.

\section{Trade-off between sensitivity and linearity} \label{sec:trade-off}

One of the most significant advantages of Fourier-type wavefront sensing is its high sensitivity. However, it comes at the expense of a limited linear range of operation~\cite{Veri04}. This trade-off has been detailed in~\cite{Fauv16} by defining the sensitivity as $$s(\phi)= \norm{I_l\left(\phi\right)}_{L^2(\R^2)}$$ and the linear range as $$d(\phi)=  \norm{I_q\left(\phi\right)}_{L^2(\R^2)}^{-1},$$ where $I_l$ represents the linear term and $I_q$ the quadratic term of the intensity~\req{i1} according to Taylor's expansion around a reference phase, e.g., a flat wavefront. The considered norms are the $L^2(\R^2)$-norms. Then, the so-called SD-factor can be introduced as
$$sd(\phi)= s(\phi) \cdot d(\phi) = \norm{I_l\left(\phi\right)}_{L^2(\R^2)}\cdot \norm{I_q\left(\phi\right)}_{L^2(\R^2)}^{-1}.$$ 
The SD-factor indicates that sensitivity and linear range are competing variables. Hence, for robust system performance, it is crucial to find a suitable trade-off between these variables. 

\bigskip \par
In order to investigate the non-linearity of Fourier-type WFSs in more detail, it is necessary to consider two sensing regimes: the linear and non-linear regime.
Fourier-type wavefront sensing in the linear regime is commonly associated with the measurement of relatively low wavefront errors, i.e. $\phi << 1$ radians. This assumption is valid when the system is being operated in closed loop. Additionally, weak atmospheric turbulence conditions or long sensing wavelengths benefit sensor linearity. A common approach to increase the linear range of Fourier-type WFSs is to include modulation in the sensor designs. For instance, a tip-tilt mirror can circularly move the electromagnetic field around the apex of the optical element, thereby improving the light distribution to all faces of the optical element~\cite{Raga96}. In general, the linearity of the sensor increases with the modulation radius. However, introducing modulation comes at the expense of diminished sensor sensitivity, and additional optical hardware is needed.

\bigskip \par

For future ELTs with diameters over 30~m, a pyramid wavefront sensor without modulation can provide the sensitivity needed for demanding AO applications such as low-flux regimes. In FSOC the goal is to be able to perform optical satellite links at any time. Hence, these AO systems must be expected to operate during the day, a time during which it is typically considered that the most challenging atmospheric turbulence conditions occur. In ophthalmic AO, it is essential to keep the overall costs of an AO system as low as possible. Unmodulated Fourier-type WFSs are less complicated and cheaper than their modulated versions, and thus more appealing in medical imaging. In conclusion, there is a high demand for very sensitive and potentially simple WFSs. Ways to overcome non-linearity errors without introducing modulation to the sensor setup are promising ways forward.

\bigskip \par

When Fourier-type WFSs are employed in non-linear regimes, studies have demonstrated a time-averaged frequency-dependent loss of sensitivity~\cite{deo_spie_2018,Korkiakoski_08}. There are several ways to mitigate such errors: various OG compensation methods without and with an additional focal plane camera in the WFS setup, adding an optical preconditioner to the WFS setup assisted by AI, or the application of non-linear control algorithms, potentially also based on AI.

The general idea of OG methods is to use the standard linear interaction matrix-based approaches for AO control, but to adapt them. These adaptions depend on OG estimations that are a set of scalar values encoding the loss of sensitivity with respect to each component of a modal basis~\cite{deo_spie_2018,Korkiakoski_08}. The OG compensation can also be done by applying the so-called specific matrix which represents the effects of self-modulation. This matrix depends on the unknown incoming wavefront. Therefore, only an approximation of it is available~\cite{Cisse2024}. To further improve the results of the OG compensation, a focal plane camera can be added to the Fourier-type WFS setup~\cite{Cham2021}. This camera provides additional information on the incoming wavefront and supports the accurate estimation of OGs.

Another hardware-based technique is inspired by artificial intelligence~\cite{Guzman_24}. Again, the WFS is controlled with a linear interaction matrix-based approach. However, the authors suggest incorporating a diffractive element into the optical path of a PWFS. This diffractive element is trained by taking into account the reconstruction method through an end-to-end scheme. The diffractive layer of the new deep PWFS acts as an optical preconditioner. The performance of the PWFS is improved by extending the linear range of operation without applying active optical modulation. 

A third way to tackle the non-linearity of Fourier-type wavefront sensing is to apply non-linear control algorithms rather than linear approaches like interaction matrix-based MVMs. Examples of such studies for the PWFS are a non-linear estimation based on Newton's method~\cite{Frazin2018}, non-linear Landweber iteration based on simplified PWFS models~\cite{HuRa18_2} and non-linear reconstructors based on CNNs~\cite{Landman2024}.

In the next section we present an accelerated non-linear solver that is straightforwardly applicable not just to the PWFS but to all Fourier-type WFSs.

\section{Non-linear solver for Fourier-type wavefront sensing} \label{sec:nonlinear_solver}

To reconstruct a wavefront we aim to solve the inverse problem
\begin{equation*}\label{n1}
 s= I(\phi)
\end{equation*}
from noisy WFS data $s^\delta$ fulfilling for some noise level $\delta > 0$
\[ \norm{s-s^\delta}_{L^2(\R^2)}\le\delta\,. \]
For Fourier-type WFS data, the intensity operator $I$ is defined according to~\req{i1}, hence, non-linear in $\phi$. This work proposes the application of non-linear iterative algorithms for wavefront reconstruction to handle the effects of non-linearity of Fourier-type wavefront sensing.

Although interaction matrix-based approaches, potentially combined with OG compensation methods, are still the standard methods for AO control, non-linear iterative wavefront reconstruction algorithms have already been investigated in~\cite{Frazin2018,HuRa18_2} for the $4$-sided PWFS. Similar to~\cite{HuRa18_2} we propose to use the non-linear Landweber iteration. However, unlike the work presented here, the algorithm was only applicable to PWFS data and the underlying mathematical models were heavily simplified. In~\cite{HuNeuSha_2023} the authors performed a thorough mathematical analysis of the underlying model for Fourier-type WFSs. Based on the mathematical derivations, they introduced a general non-linear reconstructor for all Fourier-type WFSs using Landweber iteration. The applicability of the algorithm was simulated exclusively for the PWFS. In this work, we present an acceleration of the method presented in~\cite{HuNeuSha_2023} and perform a feasibility study of the new algorithm for several Fourier-type WFSs as well as end-to-end simulations for ELT-scale instruments.

\subsection{Non-linear pyramid extension (NOPE)}
The proposed non-linear pyramid extension (NOPE) is a generalised solver for all Fourier-type WFSs. It combines non-linear Landweber iteration with Nesterov acceleration. The algorithm is directly applicable to any Fourier-type WFS by simply inputting the $OTF$ of the chosen optical element (see section~\ref{sec_intro}). Additionally, any imperfections from the manufacturing of the optical element can be accounted for in the reconstructor by incorporating them in the $OTF$.

Both, the Landweber iteration and the Nesterov acceleration have already been studied in-depth by the mathematical community with multiple applications in the field of inverse problems~\cite{Neub95,kaltenbacher2008iterative,Nes83}. Nesterov acceleration is a modification of the standard gradient descent method that incorporates a momentum term. The next step in the iterative process is determined by using a combination of the current gradient and the previous update. Please note that the noise level $\delta$ is omitted in the following for simplicity of notation. However, the method is applicable to noisy data respectively. For NOPE, we start with an initial guess $\phi_0 = \phi_{-1}$. Then, the iterative update rule is described by
\begin{gather}
\begin{aligned}\label{n2}
\psi_k &= \phi_k+\frac{k-1}{k+\alpha-1}\left(\phi_k-\phi_{k-1}\right), \\
 \phi_{k+1} &=\psi_k+\omega_kI'\kl{\psi_k}^*\kl{s-I\kl{
 \psi_k}}\,,\qquad\qquad k=0,1,2,\dots\,,
\end{aligned}
\end{gather}
where $\omega_k$ is an iteration-dependent step size and $\alpha\ge 3$. The common practice for the regularization parameter is $\alpha = 3$. For simplicity of notation, we define the residual $$r_k:=s-I\kl{
 \psi_k}.$$ The term $I'\kl{\psi_k}^*\kl{r_k}$ represents the adjoint of the Fr\'echet derivative of the intensity operator $I$ in $\psi_k$ in direction of the residual $r_k$. We suggest to use the steepest descent step size
 \begin{align*}
s_k &:= I'\kl{\psi_k}^*\kl{r_k}, \\
\omega_k &= \frac{\norm{s_k}_{L^2(\R^2)}^2}{\norm{I'\kl{\psi_k}\kl{s_k}}_{L^2(\R^2)}^2}
 \end{align*}
for the iteration-dependent step size in order to reduce the number of necessary iterations~~\cite{Neub18}.

As derived in~\cite{HuNeuSha_2023} the Fr\'echet derivative of the intensity operator $I$ and its corresponding adjoint are given by
\begin{align}
     I'(\phi)\left(r\right)&= 2\Re{\left(\overline{A(\phi)}A'(\phi)r\right)} \,,\\
     I'(\phi)^*\left(r\right)&=2\Re{\left(A'(\phi)^*(A(\phi)\cdot r)\right)}\,,
\end{align}
with
\begin{align}\label{n3}
    A(\phi) &:=F^{-1}\left(OTF\cdot F\left(\chi_\Omega e^{-i\phi}\right)\right)\, , \\
     A'(\phi)r&=  F^{-1}\left(OTF\cdot F\left(\chi_\Omega e^{-i\phi}\left(-ir\right)\right)\right)\, , \\
    A'(\phi)^*\left(r\right)&=i\chi_\Omega e^{i\phi}F^{-1}\left(\overline{OTF}\cdot F\left(r\right)\right)\,. 
\end{align}
The notation $\Re\left(\cdot\right)$ denotes the real part, $\overline{OTF}$ is the complex conjugate of $OTF$ (or of the operator $A$ respectively) and $\phi$,$r$ are real functions.

The Landweber iteration itself is known to have a rather slow convergence. Using the Nesterov acceleration, a speed increase from $\mathcal{O}(k^{-1})$ to a convergence rate of $\mathcal{O}(k^{-2})$ can be possible.

\subsection{Initial guess}
The iterative procedure of NOPE~\eqref{n2} starts with an initial guess $\phi_0$. It may include a priori knowledge of the exact solution $\phi_*$. If the system runs in closed-loop and the wavefront sensor is in its linear regime, an effective and simple starting value is a flat wavefront. Employing a linear solution, such as the result of an interaction matrix-based approach as a starting value, can have two advantages. Firstly, the non-linear method might improve the reconstruction quality obtained with the linear reconstructor. Secondly, the necessary number of iterations can be reduced, which in turn increases the efficiency of the algorithm.
Furthermore, it is possible to apply a warm restart to the system. Real-world AO systems run on the order of KHz, implying small wavefront changes between consecutive AO frames. Hence, the reconstruction of the previous frame can be a good initial guess for the current frame. The initial guess at the time step $t$, denoted by $\phi_{0,t}$, is chosen as the final
reconstruction $\phi_{rec,t-1}$ of the last time step $t-1$. With this warm restart strategy, it is expected that a very small number of iterations (2-5) is required after the first wavefront measurement. Therefore, a warm restart decreases the computational load of the iterative algorithm whilst simultaneously providing an optimised wavefront reconstruction performance~\cite{HuRa18_2}. 

Overall, the choice of the initial guess can be particularly important in non-linear regimes in order to ensure convergence with a low computational workload.

\subsection{Stopping rules}
For the Landweber iteration the number of iterations acts as regularization in case of noisy data $s^\delta$. As a stopping rule the discrepancy principle can be used, i.e. the iteration stops after $k_*$ steps defined via
$$\norm{s-I(\phi_{k_*})}\le \tau\delta < \norm{s-I(\phi_k)}, \qquad 0\le k < k_*$$ with a suitable constant $\tau > 1$. For non-linear Landweber iteration, several other possible stopping rules have been investigated~\cite{Hub22}. However, we found one main limitation of (heuristic) stopping rules in closed-loop AO: time-consuming computations that only slightly improve the results. Thus, an appealing alternative to the discrepancy principle or other stopping rules is to fix the number of NOPE iterations in advance. We have observed in simulations that the number of necessary iterations per frame in an AO loop for stable performance is highly dependent on the regime the WFS operates in. For sensing in the linear regime, a very small number of NOPE iterations, e.g., $3-5$ iterations, are sufficient for a stable AO control. This is due to the fact that the residual hitting the DM only varies slightly within consecutive AO frames. However, in non-linear regimes a larger number of iterations is necessary. As an example, for $r_{0}$ values below $17$~cm in R-band sensing, between $50-100$ iterations are needed for convergence. The observations above were made when a flat wavefront was used as a starting value.

We propose to define a rule to calculate an optimal stopping index for NOPE based on the sensing wavelength $\lambda$, the Fried parameter $r_0$ and the initial guess of the iterative algorithm. This can, for instance, be realised by lookup tables for various Fried parameters and sensing wavelengths. As the optimal stopping index depends on the AO system and atmospheric parameters, a library of simulation results will be required. These ideas will be explored in future work.

\subsection{Computational workload}
AO systems observe dynamic wavefronts and stable wavefront correction requires the operation of the AO loop in real-time. This study suggests an iteratively calculated wavefront update on every WFS frame. Thus, the efficiency of the algorithm is of great importance for the application of the method to real-world use cases. 

As we wish to send DM actuator commands to the AO system, the number of active actuators $n_a$ denotes the number of unknowns to be found. NOPE is an algorithm that follows the idea of a full-frame approach. This means that WFS measurements on the full detector image are considered. Depending on the variant of the Fourier-type WFS, many of the detector pixels will be zero because they are exposed to a relatively low level of light (see fig.~\ref{fig_otf}, bottom). Let $n\sim 4n_a$ indicate the number of nonzero detector pixels. NOPE consists of pointwise multiplications and summations which have computational complexities of $\mathcal{O}\left(n_a\right)$ or $\mathcal{O}\left(n\right)$. These are combined with Fourier transforms, which attribute a significant portion of the computational overhead. They are carried out in the focal and detector plane where full frame images of the size $n$ are considered. Hence, the Fourier transforms have a complexity of $\mathcal{O}\left(n \log n\right)$. Thus, the overall computational effort of NOPE is $\mathcal{O}\left(n \log n\right)$.

Please note that the solution of NOPE is a reconstructed wavefront, which then has to be transformed into mirror actuator commands. The effort of this additional projection step is not considered here. 

Ideally, the number of NOPE iterations per AO frame should be kept as low as possible. In an AO closed-loop operation, the changes of the residual wavefronts are relatively small. At the start of each frame, the algorithm can use previous results as an accurate initial guess. This warm restart technique suggests a reduction in the number of required iterations per frame. Additionally, it is possible to apply the solution of a linear solver as an initial guess. For the linear method, we suggest the preprocessed cumulative reconstructor with domain decomposition (PCuReD)~\cite{Shat13}. Having only a linear complexity, this method is, to our knowledge, the fastest reconstructor available for the PWFS. PCuReD can provide an accurate starting value for NOPE and decreases the number of necessary iterations per AO frame.

Even though it is believed that NOPE will be able to achieve real-time performance with dedicated software development, the authors would like to point out the main concern of this work: The research conducted has concentrated on finding an optimised solution to the inverse problem of wavefront reconstruction with nonmodulated Fourier-type wavefront sensing. Hence, the computational workload of the algorithm and its real-time implementation have not yet been prioritised.

\section{Numerical simulations\label{sec:numsim}}

The numerical simulations were performed using Octopus, an astronomical end-to-end simulation tool of the European Southern Observatory (ESO)~\cite{octopus06}. We simulated a single conjugate AO system on a $37$~m primary mirror telescope, i.e., a telescope on the scale of the ESO ELT. We used a von Karman atmospheric model with $35$ simulated layers~\cite{Sarazin1990}. The Fried parameter, $r_{0}$, defined here at 500nm, was simulated over 5 values: $23.4, 17.8, 15.7, 13.9, 9.7$~cm. High photon flux ($10000$ photons/pixel/frame) was assumed for every simulation. The DM had a total of $75\times 75$ actuators on a Fried geometry. The system was running at a frame rate of $1$~kHz with an integrator controller and $2$ frames DM delay. Detector read-out noise and background flux were considered in the end-to-end simulations. Wavefront sensing was performed with several different Fourier-type WFSs at a sensing wavelength of $\lambda=2.2$~$\mu$m (K-band) and $\lambda=0.7$~$\mu$m (R-band) to emulate linear and non-linear sensing regimes. The pupil resolution was $740\times 740$ with a sensing sampling of $74\times 74$ pixels. Modulation was not considered in the WFS setup. The simulation parameters are summarised in tbl.~\ref{table:simparam} and represent a typical ELT instrument. By default, the initial guess of NOPE was a flat wavefront.

\bigskip \par

\begin{table}[tbh!]
\centering
\begin{tabular}{|l | l |} 
\hline
\textbf{End-to-end simulation parameters} &  \\
\hline
Telescope diameter & $37$~m, no obstruction    \\
Science target & on-axis (SCAO)   \\ 
WFS & several Fourier-type WFSs   \\
Sensing band $\lambda$ & K ($2.2$~$\mu$m) \& R ($0.7$~$\mu$m)   \\
Evaluation band $\lambda_{science}$ & K ($2.2$~$\mu$m)  \\
Modulation  & $0$ $\lambda/D$  \\
Atmospheric model & von Karman   \\
Number of simulated layers & $35$   \\
Outer scale $L_0$ & $25$~m   \\
Fried parameter $r_0$ at $500$~nm & $\left[23.4, 17.8, 15.7, 13.9, 9.7\right]$~cm   \\
Number of sensor pixels & $74 \times 74$  \\
Number of active pixels & $3912$ out of $5476$  \\
Linear size of simulation grid & $740$ pixels \\
DM geometry & Fried $75 \times 75$ actuators \\
Controller &  integrator  \\
DM delay & $2$ frames  \\
Frame rate &  $1$~kHz   \\ 
Photon flux &   $10000$ photons/pixel/frame  \\
Detector read-out noise & $1$ electron/pixel \\
Background flux & $0.000321$ photons/pixel/frame \\
Time steps &  $500$  \\ 
\hline
\end{tabular}
\caption{Parameters for Octopus simulations.}
\label{table:simparam}
\end{table}

\subsection{Comparison with different Fourier-type WFSs}\label{sec:WFS_comparison}

\begin{figure}
\centering
  \includegraphics[width=0.55\textwidth]{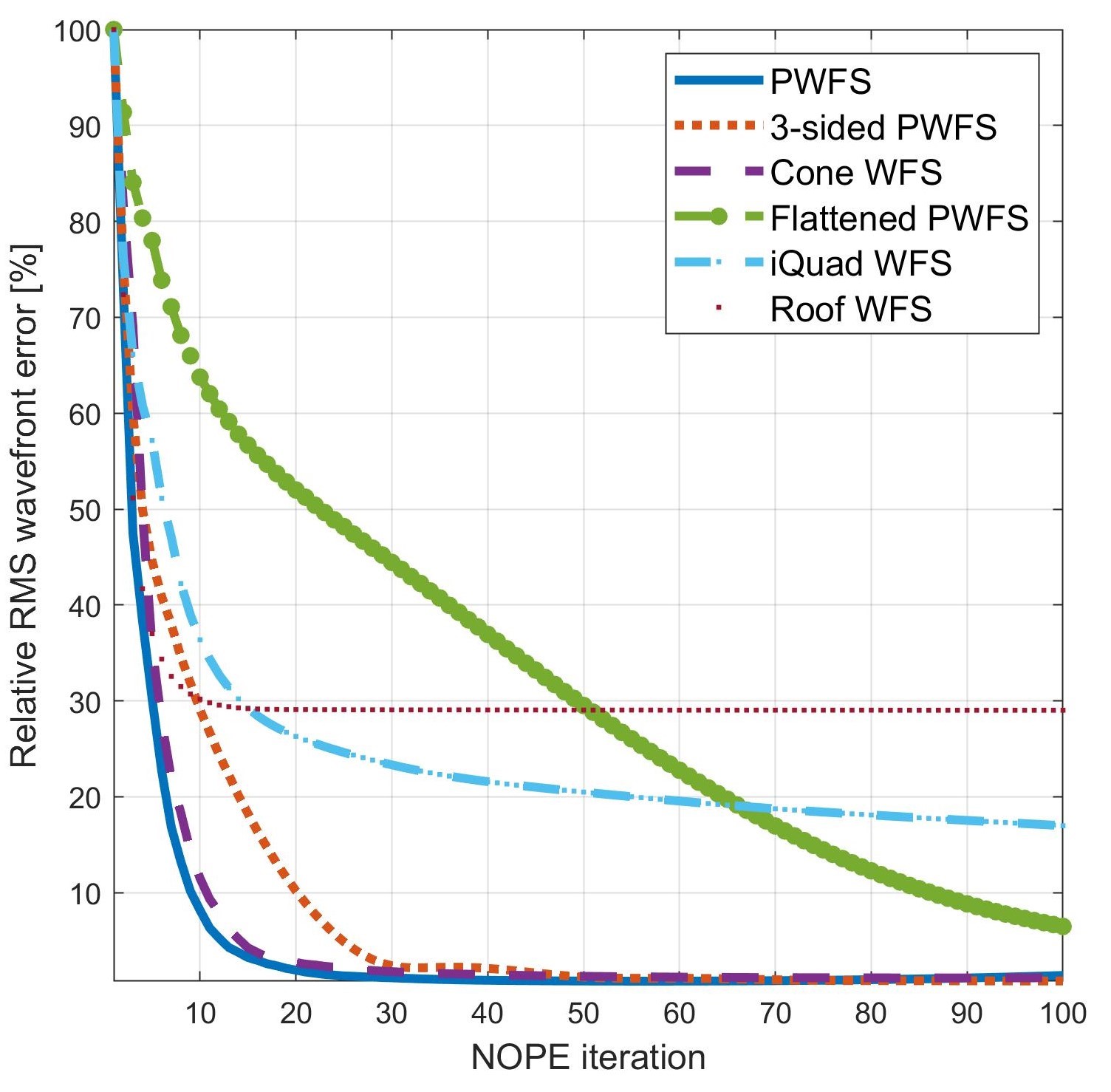}
 % \caption{Wavefront error of several Fourier-type WFSs in the linear regime.}
\caption{Wavefront error of several Fourier-type WFSs in the linear regime visualised for $100$ NOPE iterations. The relative RMS wavefront errors of the classical PWFS and the cone WFS decrease very fast. The 3-sided PWFS and especially the flattened PWFS experience a slower convergence. This is related to parameters being fine-tuned to the 4-sided PWFS only. However, the roof WFS and the iQuad WFS show a stagnating residual wavefront error suggesting unseen modes of the WFSs.}
\label{fig_comparison}
\end{figure}

\begin{figure}
  \centering
  \includegraphics[width=0.99\textwidth]{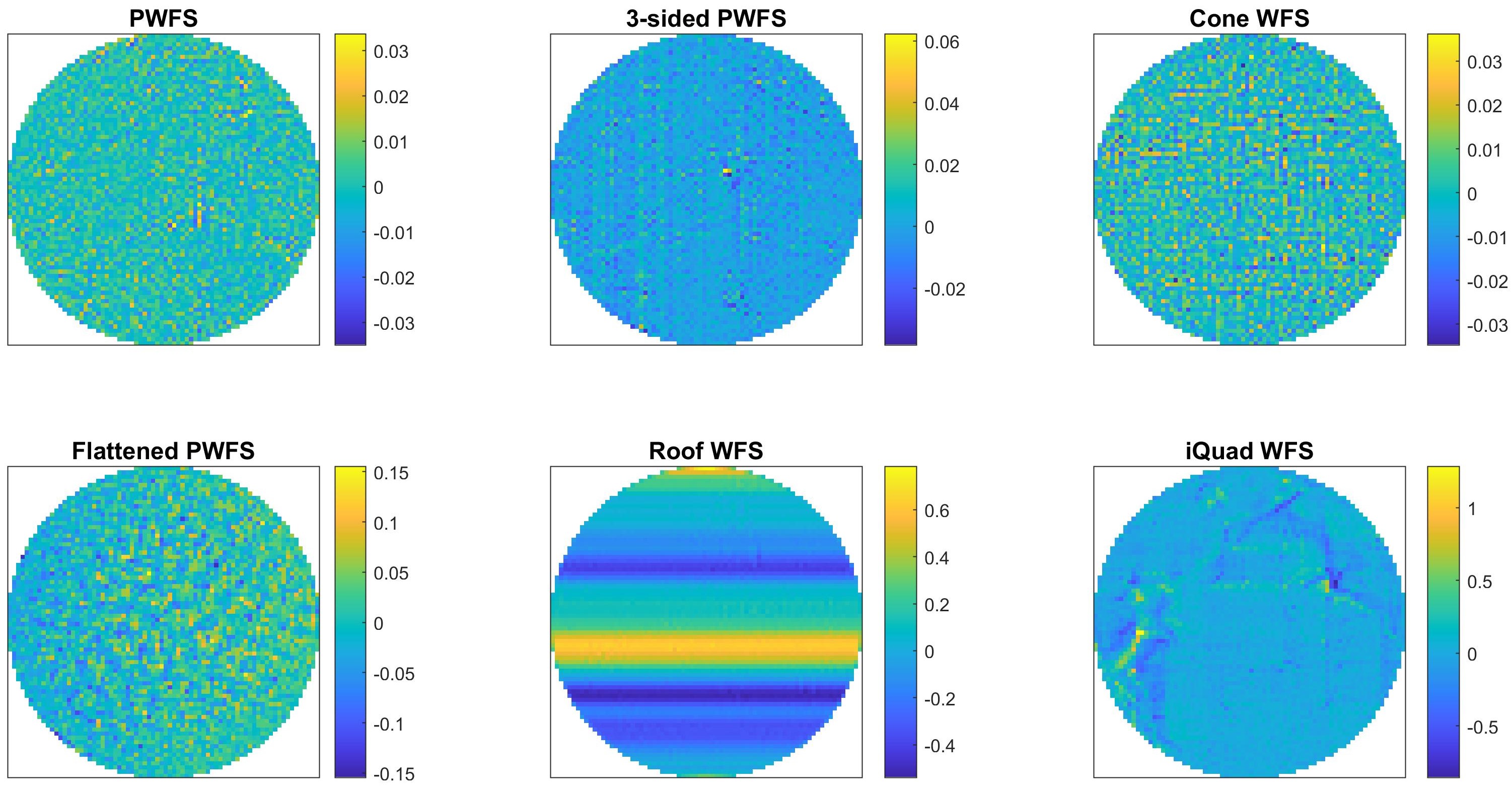}
  \caption{Residual wavefronts (in radians) potentially related to unseen modes of the WFSs. The residual wavefronts of the PWFS, 3-sided PWFS and the cone WFS correspond to very well-corrected wavefronts. The residual of the flattened PWFS suffers from slow convergence due to inadequate parameter fine-tuning. The roof and the iQuad WFS residuals show unseen modes of the WFSs.}
  \label{fig_nullspace}
\end{figure}

Firstly, we showed that NOPE could indeed be applied to different versions of Fourier-type WFSs and compared their performance. The comparison was carried out in a simulation that ignored temporal dynamics. The incoming wavefront considered for the investigations was taken from Octopus. It was scaled to ensure sensing in the linear regime. The Fourier-type WFSs considered were the classical 4-sided PWFS, the 3-sided PWFS, the flattened PWFS, the cone WFS, the roof and the iQuad WFS. Reconstruction parameters were fine-tuned for the classical PWFS and were not adjusted for other WFSs. Hence, the performance of these might improve by individual fine-tuning. Fig.~\ref{fig_comparison} shows the root-mean-squared (RMS) wavefront errors for $100$ NOPE iterations. The classical PWFS and the cone WFS provided the best performance, directly followed by the 3-sided PWFS with a similar reconstruction quality, however, slightly slower convergence. The slower convergence of the 3-sided PWFS could be related to the parameter fine-tuning. Thus, we claim exceptional reconstruction performance for all three WFSs: the 4-sided, 3-sided and the cone WFS. The convergence of the flattened PWFS is relatively slow. This is again related to the parameter fine-tuning and can be overcome by a choice customised to the flattened PWFS. However, for the roof and the iQuad WFS the wavefront error stagnates. The residual wavefront errors visualised in fig.~\ref{fig_nullspace} suggest that the error for the roof and iQuad WFS is not related to the reconstructor, but to the WFSs themselves. The residuals of the classical PWFS, the 3-sided PWFS and the cone WFS (top) show characteristics of a very well-corrected wavefront. The residual wavefront error of the flattened PWFS is related to the parameter fine-tuning that induces a slower convergence. The residuals of the roof and the iQuad WFS, however, show the unseen modes of the WFSs. For example, the residual of the roof WFS represents a tilt. This is because only one roof prism has been used in the simulation. By adding a second, orthogonally placed roof prism to the WFS setup, the unseen tilt will disappear. We also know from the literature that the iQuad sensor has at least one unseen mode~\cite{FaHuLAM_ao4elt6_poster}. However, these modes have not been identified yet. Overall, we showed that NOPE is a generalised non-linear reconstructor that can be straightforwardly applied to all Fourier-type WFSs. The only adaption required is the $OTF$.

%\begin{figure}
%\centering
%  \includegraphics[width=0.99\textwidth]{comparison_wavefront_intensity_error.jpg}
% % \caption{Wavefront error of several Fourier-type WFSs in the linear regime.}
%\caption{Wavefront error (left) and intensity error (right) of several Fourier-%type WFSs in the linear regime visualised for $100$ NOPE iterations. The relative %RMS wavefront error (left) of the classical PWFS and the cone WFS decrease very %fast. The 3-sided PWFS and especially the flattened PWFS experience a slower %convergence. This is related to parameters being fine-tuned to the PWFS only. %However, the roof WFS and the iQuad WFS show a stagnating residual wavefront %error suggesting unseen modes of the WFSs. The relative RMS intensity error %(right) is minor for all WFS verifying the accuracy of the underlying model\st{in NOPE.} {\color{green} what is intensity here? photon flux? also what is the r0 for these plots? why aren't there error bars?}}
%\label{fig_comparison}
%\end{figure}

\subsection{End-to-end simulations with the PWFS} \label{sec:4pwfs}

For the end-to-end simulations we chose one of the top WFSs of the last section: the classical 4-sided PWFS. We tested and compared three different algorithms, NOPE, a zonal MVM approach and PCuReD~\cite{Shat13} in K- and R-band sensing. The MVM method used here was an approximation of the minimum variance reconstructor using a zonal interaction matrix and Laplacian regularisation~\cite{ShaHuRa20}. Optical gain compensation was not included. The second reconstructor was the PCuReD algorithm. PCuReD is a linear reconstructor for the PWFS based on a simplified version of the underlying mathematical model. It includes a preprocessing kernel that is related to the PWFS pixel size, the modulation pattern and the sensing wavelength. For the simulations below, we chose a default PCuReD kernel. While having a computational complexity of just $\mathcal{O}\left(n_a\right)$, PCuReD was shown to provide reconstruction quality comparable to interaction-matrix-based approaches. For more details we refer the reader to~\cite{HuShaObRaLAM19_AO4ELTproc,Shat13}. For the non-linear reconstruction we chose to use the fast PCuReD or MVM with NOPE in tandem in order to save computational time. The initial guess for NOPE was again a flat wavefront. 

For optical gain compensation, individual basis functions of the wavefront are weighted differently to compensate for the respective sensitivity loss of the WFS. Frequency-dependent reconstruction has been observed to be especially important in non-linear regimes. NOPE also offers such a possibility via an adaption of the algorithm. Rather than reconstructing the complete wavefront at once, we apply the NOPE steps to separated frequency ranges (e.g. low-order and high-order modes) of the wavefront cyclically. This makes it possible to weigh the separate frequency spaces of the wavefront differently. For instance, low-order modes can be emphasised to speed up the convergence rate towards closed-loop operation.

Example simulations showed that the LE Strehl ratio of the standalone NOPE can be improved when frequency-dependent reconstruction is added. For the PCuReD-NOPE combination, a frequency-dependent control is already implemented in the PCuReD preprocessing kernel. The simulation results indicate that an additional frequency-dependent control in NOPE is not needed when both algorithms are used in tandem. Hence, the optimised choice in nonlinear regimes with respect to reconstruction quality and computational time is the PCuReD-NOPE combination. 

\bigskip\par

\begin{figure}[tb!]
\centering
\includegraphics[width=0.99\textwidth]{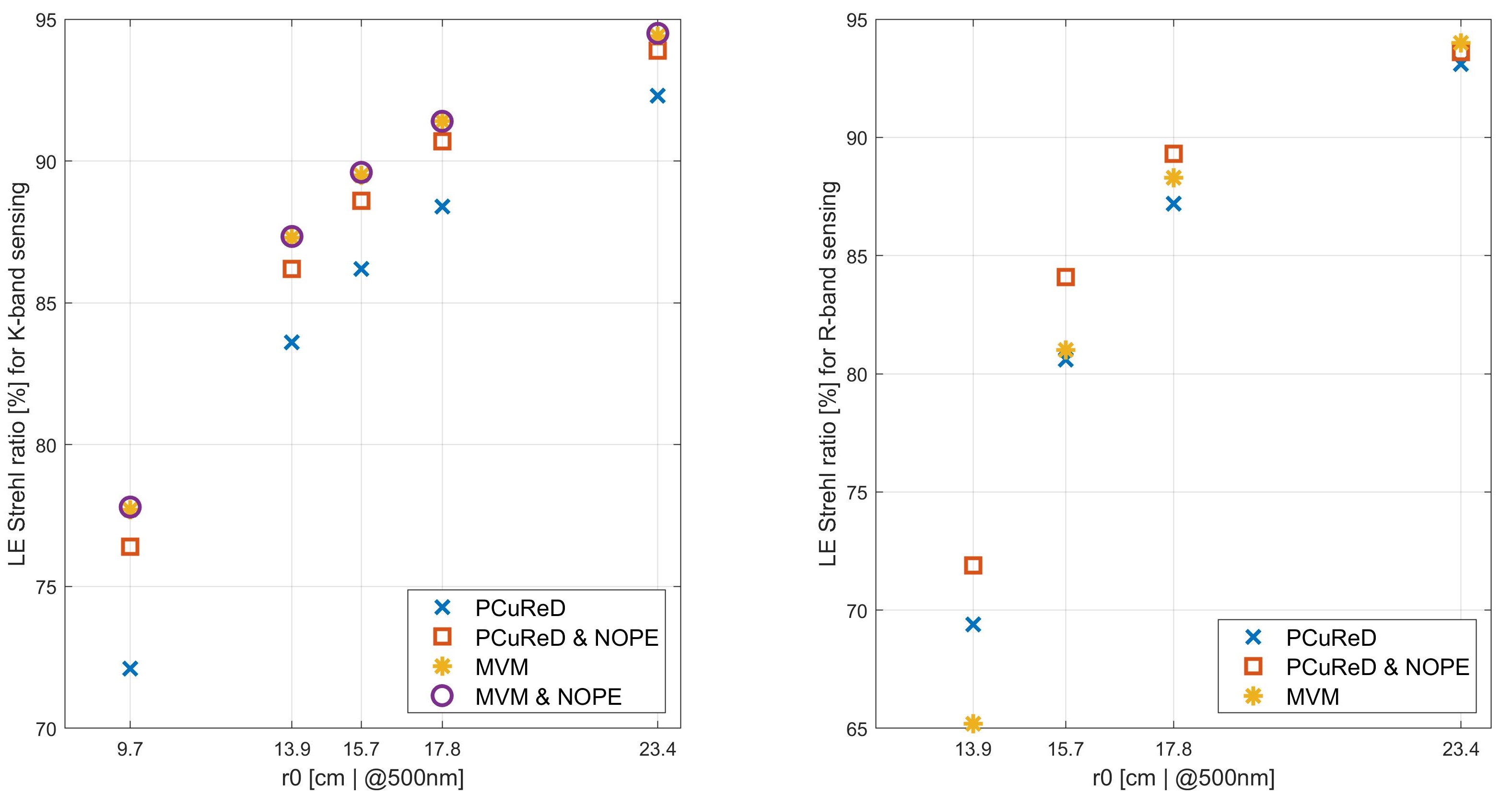}
\caption{End-to-end simulation results in Octopus at $\lambda=2200$~nm (left) and $\lambda=700$~nm (right) for combinations of linear reconstructors (MVM and PCuReD) and the non-linear NOPE. In K-band sensing the application of the non-linear reconstructor is superfluous. In R-band sensing the PWFS performance can be improved by NOPE compared to linear methods. The gain increases with decreasing $r0$. }
\label{fig_e2e}
\end{figure}

Fig.~\ref{fig_e2e} shows simulation results for $5$ different Fried parameters, $r_0$. The quality criterion is the long-exposure (LE) Strehl ratio in K-band. In K-band sensing and for $r_{0}=23.4$~cm in R-band sensing NOPE was implemented with less than $5$ iterations per AO frame. For lower $r_{0}$ in R-band the NOPE iterations increased to $50-100$. 

Fig.~\ref{fig_e2e}, left, visualises the results for K-band sensing. The standalone PCuReD performance can be improved by combining it with NOPE. However, both methods are outperformed by the MVM approach which was heavily fine-tuned. In constrast, PCuReD was run with the default kernel, and quality improvements might still be possible. The gain of the MVM-NOPE combination is about $0.1\%$ Strehl ratio compared to MVM standalone, hence negligible. Overall, the chosen MVM is a suitable reconstructor for K-band sensing with Fried parameters down to about $10$~cm. A combination with the non-linear reconstructor that adds additional computation time is superfluous. Fried parameters below $9.7$~cm were not considered.

Fig.~\ref{fig_e2e}, right, visualises the results for R-band sensing. The conclusions for very good seeing with $r_0=23.4$~cm are similar to those in K-band sensing. The application of the non-linear algorithm is not necessary. For lower $r_{0}$ we see the effects of the PWFS non-linearity. Both linear methods are now outperformed by the PCuReD-NOPE combination. With decreasing $r_{0}$ the quality improvement obtained with NOPE increases. Due to simulation time constraints, the MVM-NOPE combination was not taken into account in R-band sensing. Overall, the application of the non-linear reconstructor becomes particularly important for an $r_{0}$ below $18$~cm and linear methods are surpassed. For $r_{0}=9.7$~cm, we were unable to close the AO loop with the straightforward application of MVM or PCuReD and no OG compensation. It is understood that challenging atmospheric conditions are of particular interest for non-linear wavefront sensing and reconstruction. Future work is dedicated to investigations of NOPE for Fried parameters below $10$~cm and comparisons with MVMs that are coupled with OG compensation.

\section{Conclusions and outlook}\label{sec:conclusions}
This work presented Fourier-type wavefront sensing with a non-linear iterative algorithm for advanced AO performance. The introduced algorithm is based on the underlying mathematical model of Fourier-type WFSs. It is an implementation of the non-linear Landweber iteration with Nesterov acceleration, both well-known in the field of inverse problems. A main advantage of the NOPE reconstructor is its direct applicability to any Fourier-type WFS. The only necessary modification is the $OTF$ of the specific WFS that is being considered. 

Several Fourier-type WFSs were simulated and the suitability of the new method for different implementations of WFSs was demonstrated. Further experimental results included astronomical end-to-end simulations of ELT-like settings with a PWFS across different atmospheric turbulence conditions and sensing wavelengths. In K-band sensing with Fried parameters above $10$~cm it was shown that a heavily fine-tuned MVM provides acceptable performance, and the combination with NOPE can be circumvented to save computation time. In R-band sensing with Fried parameters below $18$~cm it was demonstrated that NOPE outperforms linear methods like MVM. Upcoming studies will determine how NOPE compares with linear methods when they are combined with OG compensation.

Future work is required to investigate the performance of NOPE during the most challenging atmospheric turbulence conditions, i.e. when $r_{0}$ is below $10$~cm. Futher simulations across varying photon counts are necessary. It is also planned to define the border between the linear and non-linear regimes of Fourier-type WFSs, i.e., to establish a rule when the application of NOPE becomes beneficial over standalone linear methods. Moreover, a thorough comparison of different Fourier-type WFSs under different atmospheric conditions will be the focus of an upcoming investigation.

\newpage
 
 \section{Acknowledgments}
This work has been funded by the Austrian Science Fund (FWF): F6805-N36. The author declares no conflict of interest. Victoria Laidlaw thanks Andreas Obereder for carrying out simulations. Victoria Laidlaw dedicates this work to her wee son Matteo.

\bibliographystyle{plain}
\bibliography{nope} 

\end{document}